\documentclass[aps,prl,showpacs,groupedaddress,
               showkeys,amsmath,amssymb,twocolumn,floatfix]{revtex4}

\usepackage{graphicx}
\usepackage{bm}
\usepackage{bbm}
\input epsf
\input rotate

\begin{document}

\title{Rheology of dilute suspensions of vesicles and red blood
cells}
\author{Victoria Vitkova$^{1,2}$, Maud-Alix Mader$^{1}$, Chaouqi Misbah$^{1}$ and Thomas Podgorski$^{1}$}
\email[]{thomas.podgorski@ujf-grenoble.fr}
\affiliation{$^{1}$Laboratoire de Spectrom\'etrie Physique--UMR 5588,
CNRS - Universit\'e Joseph Fourier, BP 87, 38402
Saint Martin d'H\`eres, France \\$^{2}$Georgi Nadjakov Institute of Solid State Physics, 72 Tzarigradsko Chausse Blvd, 1784 Sofia, Bulgaria}

\date{\today}

\begin{abstract}

We present rheology experiments on dilute solutions of vesicles and
red blood cells (RBC). Varying the viscosity ratio $\lambda$
between internal and external fluids, the microscopic dynamics of suspended objects can be qualitatively changed from tank-treading ($tt$) to tumbling ($tb$).
We find that in the $tt$ regime the viscosity
$\eta$, decreases when 
$\lambda$ increases, in contrast with droplet emulsions and elastic capsule theories which are sometimes invoked to model RBC dynamics. At a
critical $\lambda$ (close to the $tt$-$tb$ transition) $\eta$
exhibits a minimum before it increases in the $tb$ regime. This is
consistent with a recent theory for vesicles.
This points to the nontrivial fact that the cytoskeleton in RBC does not alter the qualitative evolution of $\eta$ and that, as far as rheology is concerned, vesicle models might be a better description.

\end{abstract}



\maketitle

Complex fluids are ubiquitous in nature, and
their study embraces a large spectrum of disciplines: physics, biology,
chemistry, engineering... Prominent examples in biology are
blood and cartilage. Complex fluids are also essential in several
domestic applications: cosmetic liquids (e.g. shampoo), emulsions
(e.g. mayonnaise), suspensions (e.g. clay ), etc...

Ordinary fluids and elastic solids obey universal laws, namely
Navier-Stokes and Hooke-Lam\'e laws. Contrariwise, complex fluids
still continue to escape a universal description (if any).  The
difficulty lies in the intimate coupling between the
microscopic dynamics and the macroscopic flow.
For ordinary fluids, like water, the
molecular time scale (molecular rotation, and vibration) is in the
range of $10^{-12}$~s, which is much smaller than macroscopic
flow scales, so that an adiabatic elimination of molecular modes in
favor of hydrodynamical ones 
is legitimate. In contrast, for a complex fluid like blood, red blood
cells (RBC) move and deform over time scales comparable to the global flow,
so that the macroscopic constitutive law should carry
information on the dynamics of the suspended entities, even if the
law is averaged.

The understanding of some complex features of the rheology of vesicle and RBC suspensions has recently progressed thanks to a coupling between experiments and modeling.
Our experimental study focuses on the rheology of dilute
suspensions of vesicles and RBC. The study of vesicles under
nonequilibrium conditions  keeps receiving an increasing interest. They constitute a relatively simple system (albeit complex
enough at the absolute level) that is  believed to capture some
features of RBC.

We choose to study a dilute suspension for the following reasons. (i) This
is a simple starting point which avoids additional complexity like hydrodynamic interactions between objects or rouleaux formation for RBC's 
(ii) As for other complex fluids (e.g. dilute polymer solutions where the
celebrated Oldroyd B model constitutes a basic
reference\cite{Bird87}) this offers a favorable terrain for
confrontation between experiments and modeling based on a
full microscopic description of the suspended entities. By focusing
simultaneously on vesicles and RBC, we can draw information
regarding rheology, about similarities and
differences of the two systems.

A key parameter in the present study is the viscosity contrast
$\lambda=\eta_{in}/\eta_{out}$, where $\eta_{in}$ and $\eta_{out}$
are the viscosities of the internal (hemoglobin for RBC, sugar + dextran solution for vesicles) and the
external solution (e.g. buffer + dextran, water + dextran).

Vesicles are known to exhibit three kinds of motions: (i)
tank-treading $tt$\cite{Kraus96}, in that the vesicle orients
steadily itself with a certain angle with respect to the flow
direction (while its fluid membrane makes a tank-treading motion),
(ii) tumbling when $\lambda$ exceeds a critical
value\cite{Keller82,Biben03}, and (iii)  vacillating-breathing
($vb$) mode where the vesicle long axis oscillates (or vacillates)
around the flow direction, whereas its shapes undergoes a breathing
motion( see \cite{Misbah06,Petia07,Gompper07,Danker07a,Lebedev07}
for theory, and \cite{Kantsler06,Mader06} for experiments.). An interesting question immediately arises: do the microscopic dynamics (\emph{tt}, \emph{tb} or \emph{vb}) and the transitions between them have a signature on the macroscopic level of rheology?

A major result reported here is the fact that the experimental rheology of RBC suspensions shows the same general trend as the theory for vesicles: the effective viscosity exhibits a minimum in
the vicinity of the $tt$-$tb$ transition. On the one hand, this
reveals a qualitative change due to the link between microscopic and macroscopic dynamics.
On the other hand, this suggests
that the cytoskeleton does not induce a significant qualitative change on
rheology. This behavior is consistent
with recent theoretical studies on vesicle
suspensions\cite{Danker07,Danker07a}, and contrasts with other theories sometimes invoked for RBC dynamics, like
droplet\cite{Taylor34} and capsule\cite{Barthes81} theories, where the viscosity is predicted to increase
with $\lambda$.

\begin{figure}
\centering
{
\includegraphics[width=8cm]{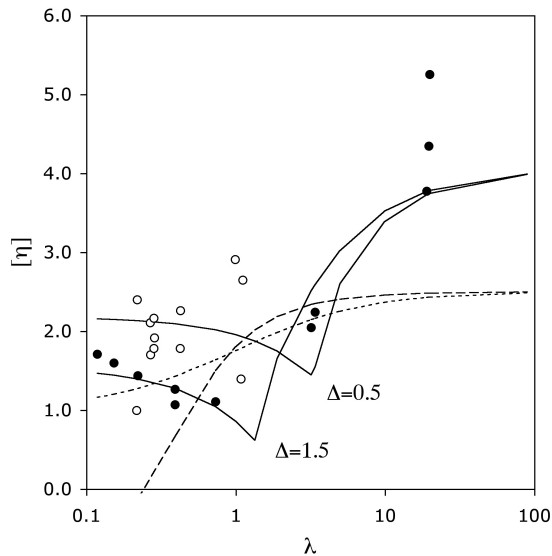} 
}
\caption{Intrinsic viscosity $[\eta]$ vs viscosity ratio $\lambda$ for RBC's with $\phi \simeq 5$\% $(\bullet)$, vesicles $(\circ)$, and various models (dotted line: drops, dashed line: capsules with $\varepsilon=0.1$ \cite{Drochon03}, plain line: vesicles with $\Delta=0.5$ and $\Delta=1.5$).
}
\label{visco_all} 
\end{figure}

{\it Theory:}
Einstein\cite{Einstein1906} provided the famous expression of
the effective viscosity of a dilute suspension of rigid spherical
particles. The intrinsic viscosity is given by
\begin{equation}
[\eta]= \frac{\eta-\eta_0}{\eta_0 \phi}=\frac{5}{2} \label{eta_rigid}
\end{equation}
where $\eta_0$ is the solvent viscosity and $\phi$ the volume fraction of particles.
Later, Taylor\cite{Taylor34} provided the analogous expression for an
emulsion: 
\begin{equation} [\eta] = \frac{5\lambda/2
+1}{\lambda +1} \label{eta_drop}
\end{equation}
A step further consists in adding a membrane around the drop. This membrane can either be an elastic solid (capsule) for which expressions for the viscosity of the suspension have been suggested \cite{Drochon03}, or a fluid lipid bilayer (vesicle). Recently an expression has been derived in the $tt$ regime for
quasi-spherical vesicles\cite{Misbah06} \begin{equation}
[\eta]_{tt}= \frac{5}{2} -\Delta \frac{23\lambda +32}{16 \pi }
\label{eta_tt}
\end{equation}
(the subscript stands for tank-treading motion), where $\Delta$ is
the excess area relative to a sphere $\Delta = (A -4\pi r^2 )/r^2$
where $r$ is the sphere radius. 
Several noticeable differences with droplets and capsules can be mentioned:
(i) the
viscosity {\it decreases} with $\lambda$ for vesicles, while the
contrary is found for droplets and capsules. (ii) When $\lambda$ is large
expression (\ref{eta_tt}) does not tend to $5/2$, as does
(\ref{eta_drop}). Expression \ref{eta_tt} is valid in the tank-treading 
regime only \cite{Misbah06}. 
At low shear
rate\cite{Danker07a} there is a direct bifurcation from $tt$ to
$tb$ when increasing $\lambda$. 
In this limit, one can assume a shape-preserving motion.
Following the general expression  for the
instantaneous viscosity \cite{Danker07}, 
$[\eta]$ can be derived in the $tb$ regime (technical details will be given elsewhere):
\begin{equation}
[\eta]_{tb}= \frac{5}{2} +\sqrt{\frac{30}{\pi}}\left[\frac{\sqrt{\Delta
-4h^2}}{\sqrt{\Delta+ 4h^2} +\sqrt{\Delta} }- h\right]
\label{eta_tb}
\end{equation}
with $h=60\sqrt{2\pi/15}/(23\lambda+32)$. The tumbling
domain corresponds to $4h^2<\Delta$ (the opposite limit is the
domain of $tt$)\cite{Misbah06}. Figure \ref{visco_all} shows the behavior of
$[\eta]$ for vesicles in both the $tt$ and $tb$ regimes. At the
bifurcation, one has a cusp singularity, with a linear behavior on
the $tt$ side, and a square root singularity on the $tb$ side.

{\it Materials and methods:}
Viscosities of diluted vesicular and red blood cell suspensions were measured as a function of the viscosity ratio $\lambda$ between the inner and outer fluids (for the vesicle or erythrocyte membrane). Chemicals were purchased from Sigma-Aldrich (France) and used without any further purification. Microscopic observations were performed on an Olympus IX 71 inverted microscope in phase contrast (GUVs and RBC) or bright field (RBC) modes. An Eppendorf 5804 centrifuge was used during sample preparation.

Erythrocyte samples were provided by the CHU (Centre Hospitalier Universitaire) of Grenoble (France) from haematologically healthy donors. Phosphate buffer saline (PBS, containing 0.01 M phosphate buffer (10.1 mM Na$_2$HPO$_4$ and 1.8 mM KH$_2$PO$_4$), 0.0027 M KCl and 0.137 M NaCl) with pH 7.4 and osmolarity $290 \pm 10$ mOsm/kg or PBS Ð dextran solutions have been used as suspending media. Blood samples were gently centrifugated at 1200 \emph{g} for 5 min and the buffy coat was removed. Erythrocytes were washed several times that way with PBS and  the hematocrit was measured via a standard automated analyser. 
The healthy biconcave shape of erythrocytes in the studied suspensions has been systematically observed in phase contrast or bright field microscopy. The cytoplasmic viscosity of RBC is determined on the basis of the mean corpuscular haemoglobin concentration (MCHC) \cite{Ross77}. At 22$^\circ$C it was estimated to be of the order of 20 mPa.s \cite{Kelemen01}. The final samples were prepared by diluting starting suspensions into a solution of dextran (D4751 and D1037 from Sigma, with respective molecular weights of and 64 000--76 000 and 425 000--575 000) in PBS buffer or pure buffer. Because of the aggregation effect of dextran on RBC \cite{Brooks73, Barshtein98, Neu02}, only dextran polymers with concentrations higher than 7\% w:v (7 g / 100 ml) were dissolved in the buffer in order to modify $\lambda$. Thus, adding up to 25\% of dextran in the suspending medium, $\lambda$ was varied in the range of 0.12-Ð3.41 and 18.37--20 without dextran.

Vesicles were prepared using the electroformation method \cite{Angelova92} with electroformation chambers designed to maximize the vesicle yield and concentration. We used DOPC, or (dioleoyl phosphatidylcholine, P-6354 from Sigma), and the electroformation solution is a sucrose solution with a concentration between 200 and 700 mM in a 1:4 glycerol-water (w:w) mixture. 
Samples were diluted (about 7 times) in an outer medium made of glucose and dextran in a 1:4 glycerol-water mixture. This outer medium is slightly hyper osmotic in order to deflate vesicles, has a different viscosity in order to vary $\lambda$ and a slightly smaller density. Diluted samples were then centrifugated (at 500--3000 rpm, corresponding to accelerations of 40--1500 \emph{g}, which gives rise to sedimentation velocities and shear rates that are safe for vesicles). 
A range of $\lambda$ between $0.2$ and $1.2$ was explored. In principle, it is possible to produce vesicles with a more viscous interior solution, and therefore reach higher values of $\lambda$ by adding dextran in the electroformation solution. However electroformation in dextran solution is much less efficient in two ways: vesicles are smaller and less concentrated 
and the dextran content is extremely heterogeneous, adding polydispersity in $\lambda$ in addition to the existing polydispersity in size and excess area of vesicle samples. The corresponding data points have therefore been discarded and are not presented in this paper. As a general rule, the preparation of  large and concentrated vesicle samples is lengthy and difficult.

Viscosity measurements were made using two rheometers: a stress-controlled Bohlin Gemini 150 rheometer (Malvern Instruments, Germany) with a cone-plate geometry (60 mm diameter, $2^\circ$ angle) and a LS30 low-shear rheometer (Contraves, Germany) with a cylinder-Couette geometry used for cases where the solvent viscosity is below 7 mPa.s. With both rheometers, the viscosity was measured across a range of shear rates between 1 and a few hundred s$^{-1}$ (for the LS rheometer: up to the upper limit of torque measurement). All viscosity measurements with GUV and RBC suspensions were performed at the constant temperature of 22$^\circ$C.

For RBC samples, solvent viscosity was measured after gently centrifugating until a completely clear supernatant is obtained. The absence of any suspended RBC has been systematically controlled microscopically. For vesicle samples, an equivalent outer medium was prepared following the same dilution procedure as the sample with pure interior solution.

Volume fractions of vesicles and RBC were estimated using different techniques. For RBC's, the final hematocrit 
was
checked by counting cells in a sample placed in a 100 $\mu$m deep PDMS observation chamber.
During the injection of RBC samples in that chamber, we could also check that   the two regimes of red blood cell dynamics ($tt$ and $tb$) were microscopically observed for $\lambda<2$ and $\lambda>2$ \cite{Goldsmith72, Fischer77}. 
 
For vesicles, after the viscosity measurement, the sample was placed in the  observation chamber (after controlled dilution to avoid vesicle overlapping on the image), and a statistical measurement of the vesicle population (vesicle diameter, number) was done using NIH Image or ImageJ software. Although this technique requires a lot of manipulation and image processing, we found that this was the most accurate way of determining vesicle volume fraction,
in contrast with centrifugation for instance.

\begin{figure}
\centering
{
\includegraphics[width=8cm]{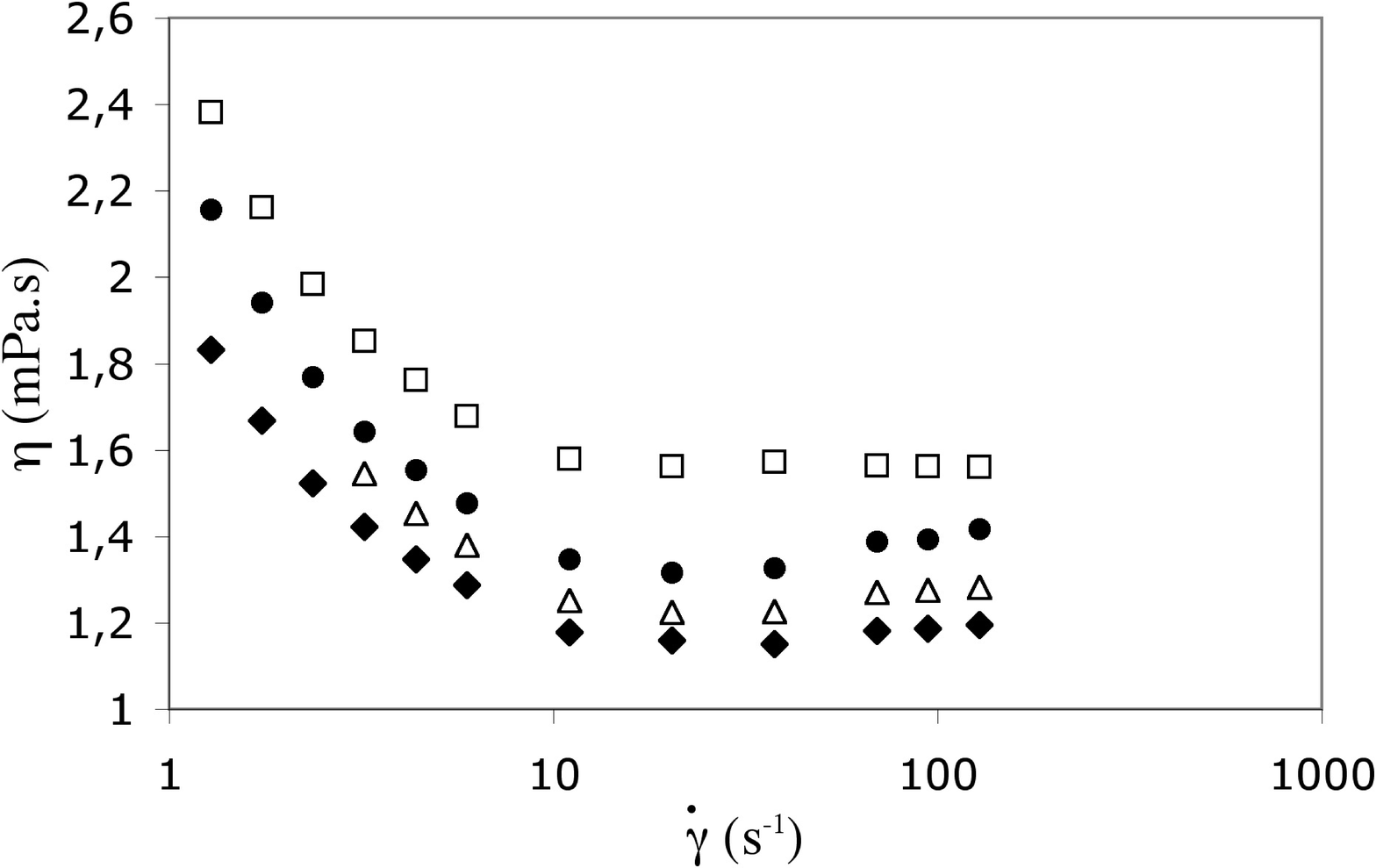} 
}
\caption{Viscosity $\eta$ vs shear rate $\dot{\gamma}$ for suspensions of red blood cells in PBS buffer ($\lambda \simeq 20$) for various volume fractions ( $(\blacklozenge)$ 3\%, $(\triangle)$ 5\%, $(\bullet)$ 7\%, $(\Box)$ 10\%).
}
\label{etavsg} 
\end{figure}

{\it Results:} 
Viscosity measurements were made in a large range of shear rates in order to detect possible viscoelastic or aggregation effects and determine the range of shear rates were measurements are accurate and correspond to the dilute suspension limit. For every sample, we retained the average value of viscosity measured in in the plateau at higher shear rates (between 10 and 100 s$^{-1}$ in the example of figure \ref{etavsg}). 
The fluctuations or different behavior at low shear rate are either due to a lack of precision of the rheometer in the corresponding torque range or to aggregation (rouleaux formation) for RBC's.

\begin{figure}
\centering
{
\includegraphics[width=8cm]{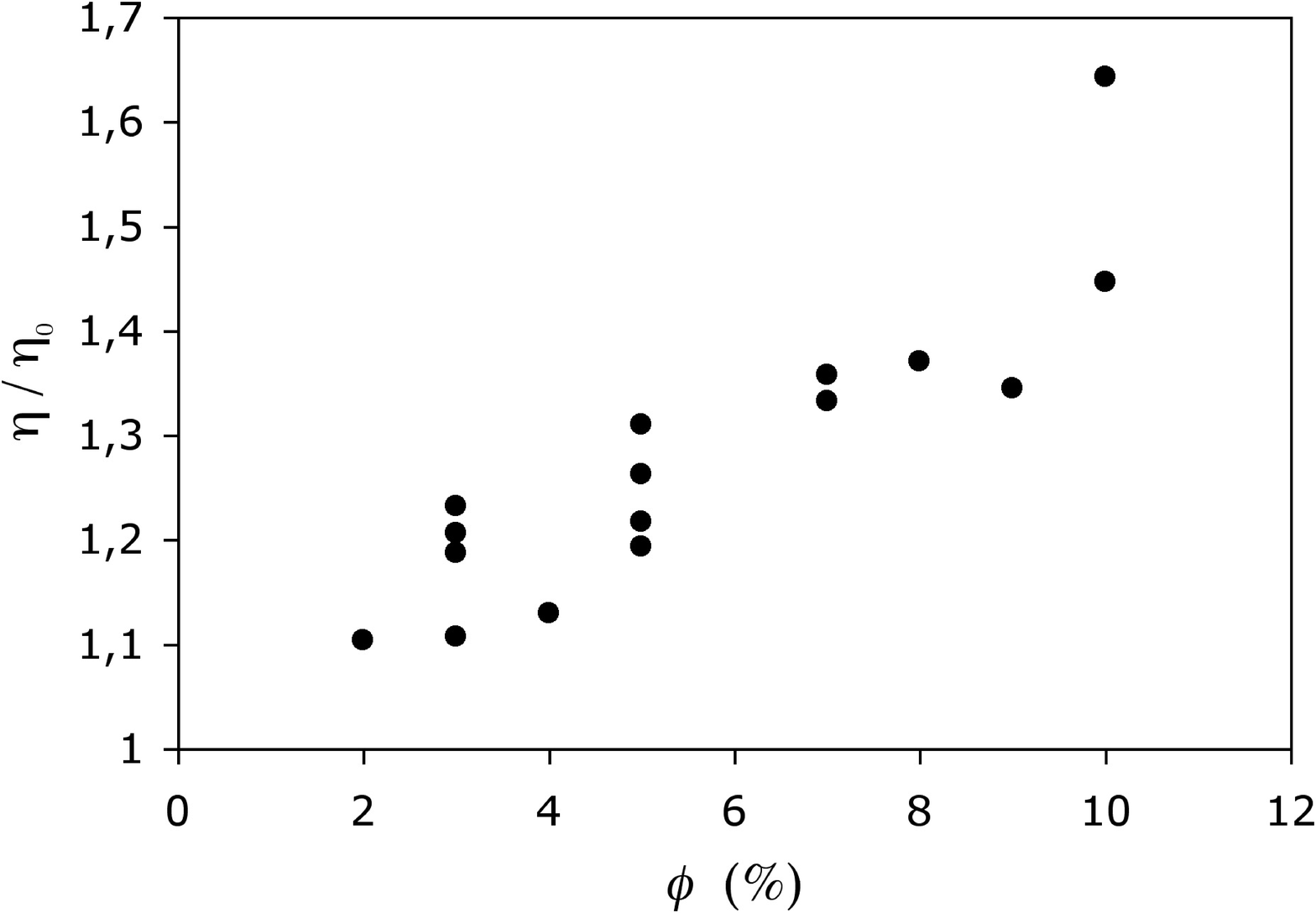} 
}
\caption{Viscosity (relative to solvent viscosity) vs volume fraction for suspensions of red blood cells in PBS buffer ($\lambda \simeq 20$), showing a linear dependency across nearly the whole range of $\phi$.
}
\label{etavsphi} 
\end{figure}

For RBC's in PBS buffer, several measurements were made at different values of the concentration (hematocrit or volume fraction) in order to check the linearity of the relationship between viscosity and concentration, and the applicability of dilute suspension theories. The results are shown in Fig. \ref{etavsphi} where up to experimental precision no nonlinear behaviour can be detected for concentrations up to 10 \%.

Figure \ref{visco_all} shows the measured intrinsic viscosities $[\eta]$ as a function of the viscosity ratio $\lambda$ for RBC's with $\phi \simeq 5$\%, and vesicles with $\phi$ ranging from 1.87 to 10.5 \%. Remarkably, the rheology of RBC suspensions follows the qualitative trends predicted by vesicle theory: a measurable decrease of $[\eta]$ in the \emph{tt} regime, a minimum near the bifurcation to \emph{tb}, and a sharp increase in the \emph{tb} regime when $\lambda$ is increased. In Fig. \ref{visco_all}, the viscosity predicted by eqs. \ref{eta_tt} and \ref{eta_tb} with $\Delta=1.5$ is shown to reproduce rather well the measured viscosity for RBC's, although their excess area is closer to $\Delta=4$. One should keep in mind however that eqs. \ref{eta_tt} and \ref{eta_tb} approximate solutions for a nearly spherical vesicle model, that does not take into account the capillary number (ratio of viscous forces and membrane bending forces). Droplet and capsule models fail to predict the correct qualitative behavior in the \emph{tt} regime and do not even approach the measured values of $[\eta]$ for high viscosity ratio, where they saturate at $[\eta]=2.5$ like suspensions of solid spheres.
For vesicles in the \emph{tt} regime, no clear tendency can be observed due to the dispersion and lack of accuracy of data points. Several factors are responsible for this: (i) there is a polydispersity of vesicle's excess area in the samples, (ii) vesicle's sizes are polydisperse, leading to different capillary numbers in the same sample, (iii) for most samples, the volume fraction is rather small (a few percent), leading to small absolute viscosity variations and amplifying errors when computing the effective viscosity. However, Fig. \ref{visco_all} reveals that the measured values of $[\eta]$ are compatible with vesicle theory \cite{Misbah06} for $\Delta=0.5$, a value consistent with measured excess areas of vesicles, and above the values obtained for RBC's, which have a larger excess area. 

Remarkably, the rheology of RBC suspension can be described by vesicle theory even though it is a simpler object, which seems to capture the essence of RBC dynamics under flow, as far as rheology is concerned. A clear transition between the  \emph{tt} and \emph{tb} regimes is observed, with a minimum effective viscosity around the bifurcation. The data obtained for vesicle suspensions is compatible with the model and confirms the predicted dependency on excess area. For a more detailed and quantitative experimental study of the rheology of vesicle suspensions, an important step forward in the production of controlled samples is a prerequisite. A decisive improvement would be the possibility to drastically reduce the polydispersity of vesicle size, excess area and viscous content.

\emph{Acknowledgements}
The authors wish to thank G. Danker for helpful and enlightening discussions, C. Verdier for experimental advice and help with rheometry, A. Drochon for helpful discussions and advice on RBC manipulation, B. Polack (CHU Grenoble) for supplying experimental material (RBC), G. Coupier for assistance on vesicle sample characterization, and E. Bayma and R. Auz\'ely for access to the low-shear rheometer at CERMAV laboratory. Financial support from CNES is gratefully acknowledged.

\end{document}